# Input Voltage and Current Sensorless Control of a Single Phase AC-DC Boost Converter

Mehdi Tavan and Kamel Sabahi

*Abstract*— It is well-known that measuring the input voltage and current, as the feedforward and feedback terms, are vital for the controller design in the problem of power factor compensation of an AC-DC boost converter. Traditional adaptive scenarios correspond to the simultaneous estimation of these variables are failed because the system dynamics is not in the classical adaptive form. In this paper, the system dynamics is *immersed* to a proper form by a new filtered transformation to overcome the obstacle. The phase and amplitude of the input voltage along with the input current is estimated from the output voltage with global convergent. An application of the proposed estimator is presented in conjunction with a well-known dynamic controller. The estimator and controller performances are evaluated by some realistic simulations.

## I. INTRODUCTION

Consider the single-phase full-bridge boost converter shown in Fig. 1. The circuit of the converter combines two pair of transistor-diode switches in two legs to form a bidirectional operation. The switches in each leg operate in complementary way and are controlled by a PWM circuit. The dynamic equations describing the average behavior of the converter can be obtained using the Kirchhoff's laws as

$$L\frac{di}{dt} = -uv + v_i(t), \quad (1)$$

$$C\frac{dv}{dt} = ui - Gv, \quad (2)$$

where $i \in \mathbb{R}$ describes the current flows in the inductance $L$, and $v \in \mathbb{R}_{>0}$ is the voltage across both the capacitance $C$ and the load conductance $G$. The continuous signal $u \in [-1, 1]$ operates as a control input and is fed to the PWM circuit to generate the sequence of switching positions $\delta_1$ and $\delta_2$ their complements $\bar{\delta}_1$ and $\bar{\delta}_2$, respectively. The switch position function takes the values in finite set $\{0, 1\}$ and the exact model can be obtain by standing $\delta_1 - \delta_2$ as the control signal in (1), (2). Finally, $v_i(t) = E\sin(\omega t + \rho)$ represents the voltage of the AC-input.

For the system, the control objective is to regulate in average the output (capacitor) voltage in some constant desired value $V_d > E$ with a nearly unity power factor in input side. This end has widely been studied by many researchers due to its applications for interfacing renewable energy sources to hybrid microgrids [1], flexible AC transmission systems [2], motor drive systems [3], LED drive systems [4] etc.

M. Tavan is with the Department of Electrical Engineering, Mahmudabad Branch, Islamic Azad University, Mahmudabad, Iran (e-mail: m.tavan@srbiau.ac.ir).
K. Sabahi is with the Department of Electrical Engineering, Mamaghan Branch, Islamic Azad University, Mamaghan, Iran.

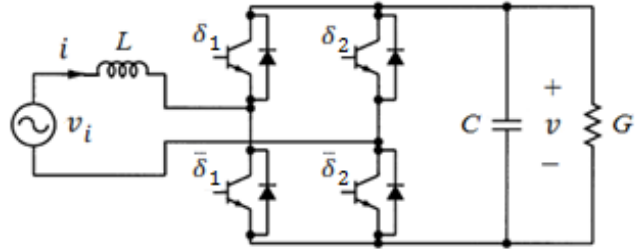

Figure 1. AC-DC full-bridge boost converter circuit [5].

Knowledge about the phase and amplitude of AC-input is vital to provide the feedforward control signal in the problem of power factor compensation (PFC) of an AC-DC boost converter; see control signals in [6] and [7]. Besides an accurate measuring of these parameters, feedback from input current improves output voltage regulation, total harmonic distortion (THD) and power factor quality [6].

It is worth noting that, the AC-DC boost converter belongs to second-order bilinear systems group and both of its states are not observable when the control signal is zero. These features pose an interesting state and parameter estimating problem which also bring down the number of sensors and costs in industrial applications. Since, an asymptotic and classical current observer is designed for the interleaved and non-interleaved AC-DC boost converters in [3, 4, 8]. Indeed, a sliding mode observer is proposed in [9] with exponential current estimation for the three phase topology of the system (1), (2). All the current observers proposed in [3, 4, 8] and [9] require a voltage sensor in input-side to measure the amplitude and phase of the AC-source, addition to the knowledge about the load conductance.

In this paper the effect of the phase difference between the input voltage and current on the power quality is investigated. The mathematic analyses show that the big values of the phase shift let to increase the DC error and the amplitude of the harmonic in the output voltage. Hence, the problem of phase difference estimation is interested in this paper. A new nonlinear, globally convergent, and robust estimator is designed via immersion and invariance (I&I) based filtered transformation. The input current and voltage are globally estimated from the output voltage via a fifth-dimensional estimator. An application of the estimator is presented in conjunction with a well-known dynamic controller. Finally, the effectiveness and robustness of the proposed scheme is evaluated using simulation results. Addition to system parasitic resistances, variation in amplitude and phase of AC-input, desired output and load conductance are considered for carring out realistic simulations.

## II. PROBLEM STATEMENT

The control objective can be achieved indirectly by means of stabilizing input (inductor) current in phase with the input voltage. This is done due to unstable zero dynamic with respect to the output to be regulated which imposed a second-order harmonic on the output in steady-state [5].

In this section we are interested to deal with the effect of phase difference between the input voltage and current on the power quality, such as the mean value of the output voltage and the harmonic wave form therein. To this end, let us to assume that the input current has the following steady-state form

$$i_s(t) = I_s \sin(\omega t + \rho - \Delta\rho), \quad (3)$$

for some constant $\Delta\rho \in (-\pi/2, \pi/2)$ as the phase difference between input voltage and current, and some $I_s > 0$ yet to be specified. Replacing (3) into (1) yields

$$u_s v_s = E \sin(\omega t + \rho) - L\omega I_s \cos(\omega t + \rho - \Delta\rho), \quad (4)$$

where $u_s$ and $v_s$ are the control input and output voltage in steady-state, respectively. Now, replacing (3) and (4) into (2) yields the following dynamics in steady-state

$$C v_s \dot{v}_s + G v_s^2 = E I_s \sin(\omega t + \rho) \sin(\omega t + \rho - \Delta\rho) \\ - \frac{L\omega I_s^2}{2} \cos(2\omega t + 2\rho - 2\Delta\rho). \quad (5)$$

The steady-state solution of (4) can be calculated using Fourier series and is given by

$$v_s^2(t) = \frac{E I_s}{2G} \cos(\Delta\rho) - \frac{G d_1 - C \omega d_2}{G^2 + C^2 \omega^2} \cos(\omega t + \rho) \\ - \frac{G d_2 + C \omega d_1}{G^2 + C^2 \omega^2} \sin(\omega t + \rho), \quad (6)$$

with

$$d_1(\Delta\rho) := \frac{I_s}{2} (E \cos(\Delta\rho) - L\omega I_s \sin(2\Delta\rho)), \quad (7)$$

$$d_2(\Delta\rho) := \frac{I_s}{2} (E \sin(\Delta\rho) + L\omega I_s \cos(2\Delta\rho)). \quad (8)$$

Finally, doing a basic trigonometric simplification (6) takes the form

$$v_s^2(t) = \frac{E I_s}{2G} \cos(\Delta\rho) + A \sin(2\omega t + 2\rho + \varepsilon), \quad (9)$$

with

$$A(\Delta\rho) := \sqrt{\frac{d_1^2 + d_2^2}{G^2 + C^2 \omega^2}}, \quad (10)$$

$$\varepsilon(\Delta\rho) := \arctan2\left(\frac{G d_1 - C\omega d_2}{G d_2 + C\omega d_1}\right). \quad (11)$$

The new representation, given by (9), demonstrates the relationship between the phase difference and the steady-state amplitude of the input current and the second-order harmonic on the output.

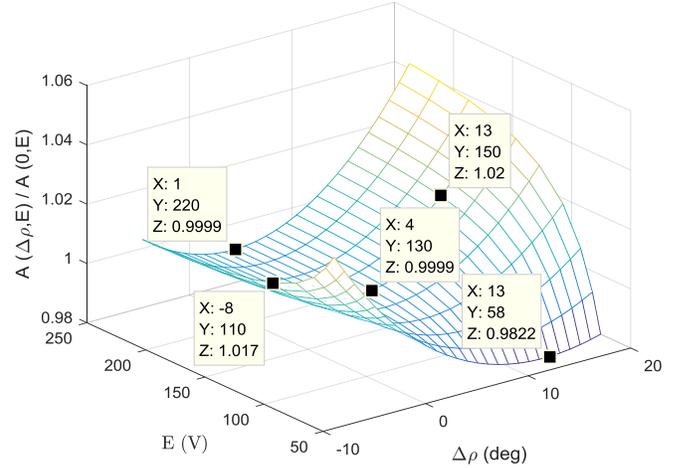

Figure 2. The influence of $\Delta\rho$ and $E$ on the amplitude of the second-order harmonic.

*Remark 1:* The DC-component of $v_s$ in steady state can be concluded from (9) and is given by

$$V_s(\Delta\rho, I_s) = \sqrt{\frac{E I_s}{2G} \cos(\Delta\rho)}. \quad (12)$$

As a result, in order to provide the desired output voltage $V_d$, the input current amplitude must be forced to achieve

$$I_s(\Delta\rho) = \frac{2G V_d^2}{E \cos(\Delta\rho)}, \quad (13)$$

which justifies the fact that for any phase shift (positive or negative) the amplitude of input current and then the converter losses increases. For $\Delta\rho = 0$, the minimum value of $I_s$ can be achieved and is defined as

$$I_0 := \frac{2G V_d^2}{E}. \quad (14)$$

As an example, consider the static control law in [5] which causes the following input current in steady-state

$$i_s(t) = \frac{I_0}{\sqrt{1 + \hbar^2}} \sin(\omega t + \rho - \arctan(\hbar)), \quad (15)$$

with $\hbar = L\omega/KE$ for some constant $K \in \mathbb{R}_{>0}$. Replacing the amplitude and phase of the current in (12) yields to

$$V_s = \frac{V_d}{\sqrt{1 + \hbar^2}}, \quad (16)$$

which indicates a steady state error in output voltage.

*Remark 2:* The minimum amplitude of the second-order harmonic $A(\Delta\rho)$ does not occur for $\Delta\rho = 0$. From (10) it can be computed that for $\Delta\rho$ which satisfies the following inequality

$$\frac{\sin^2(\Delta\rho)}{E L \omega I_0} < \frac{\sin(2\Delta\rho)(2\cos(\Delta\rho) - \cos(2\Delta\rho))}{E^2 \cos^2(\Delta\rho) + L^2 \omega^2 I_0^2 (1 + \cos^2(\Delta\rho))}, \quad (17)$$

the magnitude of $A(\Delta\rho)$ is less than $A(0)$. From (17) it can be concluded that when the converter operates in leading form, i.e. $\Delta\rho < 0$, the magnitude of $A(\Delta\rho)$ increases, but, it can be decreased when the converter operates in lagging form with a small $\Delta\rho$ near to zero.

For instance consider the experimental system used in [5] with the set of parameters shown in Table I. The ratio of $A(\Delta\rho, E)$ to $A(0, E)$ is shown in Fig. 2 for $-\cos^{-1}(0.99) < \Delta\rho < \cos^{-1}(0.95)$ and $58 \text{ V} \leq E \leq 220 \text{ V}$. Fig. 2 justifies the above statements and shows that a small lagging phase shift can be beneficial to reduce the harmonic amplitude. This may be induced by time delays or unmodeled dynamics in the actual system. On the other hand, a big phase difference, in leading or lagging form, causes high harmonic amplitude as shown in Fig. 2.

With respect to the remarks above which cite the importance of phase difference in the power quality, in this paper we are interested to estimate the parameter vector

$$\theta := [\theta_1, \theta_2]^\top$$
$$=: [E \sin(\rho), E \cos(\rho)]^\top, \quad (18)$$

from the output voltage $v$. The phase $\rho$ and the amplitude $E$ can be obtained from the above vector as

$$\rho = \arctan2(\theta_1, \theta_2), \quad (19)$$

$$E = \sqrt{\theta_1^2 + \theta_2^2}, \quad (20)$$

and the correspond regressor can be defined as

$$\varphi^\top(t) := \frac{1}{L}[\cos(\omega t), \sin(\omega t)], \quad (21)$$

to get

$$v_i(t) = L\varphi^\top \theta$$
$$= E \sin(\omega t + \rho). \quad (22)$$

## III. ESTIMATOR

In this section an estimator is designed for the system (1), (2) which requires the knowledge about $\omega$, $C$, $L$, and $G$. It should be noted that the only measurable state is the output voltage $v$. This cancels the use for input current and voltage sensor and makes the practical implementation of the proposed procedure attractive. In the design procedure the following assumptions are considered.

*Assumption 1:* The system (1), (2) is forward complete, i.e. trajectories exist for all $t \in \mathbb{R}_+$.

*Assumption 2:* The time derivative of control signal $u$ is bounded and available.

*Assumption 3:* The control signal $u$ is non-square-integrable, i.e. $u \notin \mathcal{L}_2$.

*Assumption 4:* There exists a constant $T \in \mathbb{R}_{>0}$ such that

$$\int_t^{t+T} u^2 d\tau > 0, \quad (23)$$

which implies $u$ is a persistently exciting (PE) signal, i.e. $u \in \text{PE}$.

The first assumption is standard in the estimator design and is extremely milder compared to the boundedness of trajectories. Although Assumption 2 seems to be somewhat restrictive, a dynamic control law, like the one introduced in Proposition 8.9 of [6], satisfies this assumption. Assumption 3 is needed to make the input current $i$ observable from the output voltage $v$. Moreover, Assumption 4 is required for parameter convergence. It is worth pointing out that the two previous assumptions are satisfied in operation mode when the system is forced to track a positive constant as the desired output.

The following proposition represents the main achievement of the paper.

*Proposition 1:* Consider the system (1), (2) verifying Assumption 1-2. Define the fifth-dimensional estimator

$$\begin{bmatrix}\dot{\zeta}_1\\\dot{\zeta}_2\end{bmatrix} = -k\left(\frac{u}{C}\right)^2 \begin{bmatrix}1 & -\mu^\top\mathcal{D}\\\mathcal{T}\mathcal{D}\mu & \mathcal{T}\mathcal{D}\mu\mu^\top\end{bmatrix}\left(\begin{bmatrix}\zeta_1\\\zeta_2\end{bmatrix} + \frac{u}{C}kv\begin{bmatrix}1\\\mathcal{T}\mathcal{D}\mu\end{bmatrix}\right)$$
$$-\begin{bmatrix}\frac{u}{L}\\\sigma_2\end{bmatrix}v + \frac{k}{C}v\begin{bmatrix}1\\\mathcal{T}\mathcal{D}\mu\end{bmatrix}\left(\frac{G}{C}u - \dot{u}\right) - \frac{u}{C}kv\begin{bmatrix}0\\\mathcal{T}\mathcal{D}\dot{\mu}\end{bmatrix}, \quad (24)$$

$$\dot{\mu} = -\left(\frac{u}{C}\right)^2 k(\mathcal{I}_2 + \mathcal{D})\mu + \frac{1}{L}\begin{bmatrix}\cos(\omega t)\\\sin(\omega t)\end{bmatrix}, \quad (25)$$

where $\sigma_2 \in \mathbb{R}^2$ is the zero vector, $\mathcal{I}_2$ is the $2 \times 2$ identity matrix, $\mathcal{T} = \mathcal{T}^\top$ and $\mathcal{D} = \mathcal{D}^\top$ are constant and positive definite matrixes, and $k \in \mathbb{R}_{>0}$ is a constant value.

I. If Assumption 3 holds, then

$$\lim_{t\to\infty} \bar{\iota} = 0, \quad (26)$$

$$\bar{\theta} \in \mathcal{L}_\infty, \quad (27)$$

where $\bar{\iota}$ and $\bar{\theta}$ are given by

$$\bar{\iota} := i - \hat{\iota}$$
$$=: i - [1 \quad \mu^\top]\left(\begin{bmatrix}\zeta_1\\\zeta_2\end{bmatrix} + \frac{u}{C}kv\begin{bmatrix}1\\\mathcal{T}\mathcal{D}\mu\end{bmatrix}\right), \quad (28)$$

$$\bar{\theta} := \theta - \hat{\theta}$$
$$=: \theta - \left(\zeta_2 + \frac{u}{C}kv\mathcal{T}\mathcal{D}\mu\right). \quad (29)$$

II. If Assumption 4 holds, then

$$\lim_{t\to\infty} e^{\ell t}\bar{\iota} = 0, \quad (30)$$

$$\lim_{t\to\infty} e^{\ell t}\bar{\theta} = 0, \quad (31)$$

for some constant $\ell \in \mathbb{R}_{>0}$, and

$$\lim_{t\to\infty}(\rho - \hat{\rho}) = 0, \quad (32)$$

$$\lim_{t\to\infty}(E - \hat{E}) = 0, \quad (33)$$

where

$$\hat{\rho} := \arctan2(\hat{\theta}_1, \hat{\theta}_2), \quad (34)$$

$$\hat{E} := \sqrt{\hat{\theta}_1^2 + \hat{\theta}_2^2}. \quad (35)$$

*Proof 1:* The gist of the estimator design is to change of coordinate by the parameter dependent transformation

$$\iota(t) := i(t) - \mu^\top(t)\theta, \quad (36)$$

to form a proper adaptive structure. The equation above admits the global inverse

$$i = \iota + \mu^\top \theta = \begin{bmatrix} 1 & \mu^\top \end{bmatrix} \eta, \tag{37}$$

where $\eta := \text{col}(\iota, \theta)$ is the new unavailable vector. Now, the dynamics of the system (1), (2) can be represented in terms of the new variables as

$$\dot{\eta} = k\left(\frac{u}{c}\right)^2 \begin{bmatrix} 0 & \mu^\top(\mathcal{J}_2 + \mathcal{D}) \\ \sigma_2 & O_2 \end{bmatrix} \eta - \begin{bmatrix} \frac{u}{L} \\ \sigma_2 \end{bmatrix} v, \tag{38}$$

$$\dot{v} = \frac{u}{C} \begin{bmatrix} 1 & \mu^\top \end{bmatrix} \eta - \frac{G}{C} v, \tag{39}$$

where $O_2 \in \mathbb{R}^{2\times 2}$ is the zero matrix, and use has been made of (36), (37) and (25).

Now, under the inspiration of the I&I technique, let us to define the estimation error as

$$\bar{\eta} = \eta - \zeta - \frac{u}{C} kv \begin{bmatrix} 1 \\ \mathcal{J}\mathcal{D}\mu \end{bmatrix}, \tag{40}$$

where $\zeta = \text{col}(\zeta_1, \zeta_2)$ is the estimator state. Recalling (37) in the current estimation error (28) we get

$$\bar{i} = \begin{bmatrix} 1 & \mu^\top \end{bmatrix} \bar{\eta} = \bar{\iota} + \mu^\top \bar{\theta}, \tag{41}$$

where $\bar{\iota} = \iota - \zeta_1 - \frac{u}{C} kv$, and $\bar{\theta}$ given by (29) make $\bar{\eta} = \text{col}(\bar{\iota}, \bar{\theta})$. Differentiating (40) and recalling (24), (25), (38), and (39) therein, yields

$$\dot{\bar{\eta}} = -k\left(\frac{u}{c}\right)^2 \begin{bmatrix} 1 & -\mu^\top \mathcal{D} \\ \mathcal{J}\mathcal{D}\mu & \mathcal{J}\mathcal{D}\mu\mu^\top \end{bmatrix} \bar{\eta}. \tag{42}$$

Evaluating the time derivate of the Lyapunov function

$$V(\bar{\eta}) = \frac{1}{2}(\bar{\iota}^2 + \bar{\theta}^\top \mathcal{J}^{-1} \bar{\theta}), \tag{43}$$

along the trajectories of (42) satisfies

$$\dot{V} = -k\left(\frac{u}{c}\right)^2 (\bar{\iota}^2 + \bar{\theta}^\top \mathcal{D}\mu\mu^\top \bar{\theta}) \leq -k\left(\frac{u}{c}\right)^2 (\bar{\iota}^2 + \ell_0 (\mu^\top \bar{\theta})^2), \tag{44}$$

for some constant $\ell_0 \in \mathbb{R}_{>0}$. Recalling Assumption 3 in (44) implies that the system (42) has a uniformly globally stable equilibrium at the origin, $\bar{\iota} \in \mathcal{L}_\infty$ and $\bar{\theta} \in \mathcal{L}_\infty$ then $\bar{\eta} \in \mathcal{L}_\infty$, and finally $\bar{\iota} \in \mathcal{L}_2$ and $\mu^\top \bar{\theta} \in \mathcal{L}_2$ then $\bar{i} \in \mathcal{L}_2$. Note now that (25) is an asymptotically stable filter which is perturbed by $\varphi$ given in (21). Due to $\varphi \in \mathcal{L}_\infty$ we get $\mu \in \mathcal{L}_\infty$ and $\dot{\mu} \in \mathcal{L}_\infty$. Hence, with respect to Assumption 2 and (42) it can be concluded that $\dot{\bar{\eta}} \in \mathcal{L}_\infty$ and then $\ddot{V} \in \mathcal{L}_\infty$. Therefore, the uniformly convergence of $\bar{\iota}$, $\mu^\top \bar{\theta}$ and then $\bar{i}$ to zero follows directly by Lemma 1 in [10]. This completes the proof of the point I.

The proof of the point II is established by contradiction. Assume that there exists a time instance $t_1 \in \mathbb{R}_+$ such that

$$\|\bar{\eta}\| \geq \ell_1, \quad \forall t \geq t_1, \tag{45}$$

for any constant $\ell_1 \in \mathbb{R}_{>0}$. By virtue of the assumption above, using (45) and (43) in (44) and applying some basic bounding, results in

$$\dot{V} \leq -k\left(\frac{u}{c}\right)^2 \frac{\bar{\iota}^2 + \ell_0 (\mu^\top \bar{\theta})^2}{\|\bar{\eta}\|^2} \|\bar{\eta}\|^2 \leq -\ell_2 \frac{\bar{\eta}^\top}{\|\bar{\eta}\|} Q(t) \frac{\bar{\eta}}{\|\bar{\eta}\|} V, \tag{46}$$

for some constant $\ell_2 \in \mathbb{R}_{>0}$ and

$$Q(t) = \text{diag}(u^2, u^2 \mu \mu^\top). \tag{47}$$

Integrating (46) when $t \geq t_1$ yields the solution

$$V(t) \leq V(t_1) e^{-\ell_0 \ell_2 \int_{t_1}^{t} \frac{\bar{\eta}^\top}{\|\bar{\eta}\|} Q(\tau) \frac{\bar{\eta}}{\|\bar{\eta}\|} d\tau} \leq \ell_1^2 e^{-\ell_0 \ell_2 \int_{t_1}^{t} \epsilon_0^\top Q(\tau) \epsilon_0 d\tau}. \tag{48}$$

with $\epsilon_0 := \bar{\eta}/\|\bar{\eta}\|$ which it is well-known that $\|\epsilon_0\| = 1$. The integral term in the equation above reminds us the scalar description of the PE condition. Hypothesis the condition equivalently implies the existence of a constant $T_0 \in \mathbb{R}_{>0}$ such that

$$\infty > \int_{t_1}^{t_1 + T_0} Q(\tau) d\tau > 0, \tag{49}$$

for all $t_1 \in \mathbb{R}_+$. Replacing (47) in (49) and recalling (23) from Assumption 4, ensures (49) if and only if $u\mu^\top \in \text{PE}$. Note now that, invoking the preservation of the PE property under filtering by an asymptotically stable filter [11], the condition $u\mu^\top \in \text{PE}$ can be checked for the regressor vector $u\varphi^\top$ instead. Thanks to Proposition 2 and Eq. (23) in [11], it can be concluded that

$$\infty > \int_{t_1}^{t_1 + T_0} u^2(\tau) \varphi(\tau) \varphi^\top(\tau) d\tau > 0. \tag{50}$$

Therefore from (49) and (48), for some constants $T_0$ and $\ell_3 \in \mathbb{R}_{>0}$ we obtain the bound

$$V(t) \leq \ell_1^2 e^{-\ell_3}, \quad \forall t \geq t_1, \tag{51}$$

which, in turn, implies

$$\|\bar{\eta}\| < \ell_1, \quad \forall t \geq t_1, \tag{52}$$

in contradiction with (45). As a result, $\bar{\eta} = 0$ is a uniformly attractive equilibrium point. On the other hand, $u \in \text{PE}$ induces $u \notin \mathcal{L}_2$ and from the proof of point I we get that the equilibrium point is uniformly globally stable. Hence, $\bar{\eta} = 0$ is uniformly globally asymptotically stable. Note now that, thanks to linear form of the time-varying system (42), the convergence property is exponential by Theorem B.1.3 in [12]. □

## IV. ADAPTIVE OUTPUT FEEDBACK CONTROL

The PFC main function is to achieve unity power factor. This happens if the input current to be in the phase with the AC-input. This can be achieved by the following dynamic control law [5]

$$\dot{u} = \frac{1}{v}\left(-\frac{u^2}{C}i + w\right), \quad (53)$$

where $w(t)$ is derived from the filter

$$w(s) = c\frac{s^2 + as + b}{s^2 + \omega}e(s), \quad (54)$$

with some positive constants $a$, $b$, $c$, and

$$e := v_i - L\frac{di_d}{dt} - k(i_d - i) - uv, \quad (55)$$

with positive constant value $k$, where $i_d = I_0 \sin(\omega t + \rho)$ and $I_0$ is given by (14). For sufficiently large $c$, the closed loop system is asymptotically stable, and then $e$ converges to zero and $i$ converges to $i_d$ [6]. On the other hand, from the analysis in Remark 1, for $I_s = I_0$ in (12) with zero phase difference, i.e. $\Delta\rho = 0$, the output voltage $v$ converges in average to $V_d$. Note that, $v_i$ and $i_d$ in (55) are sinusoid signals, hence we can conclude that $u$ is sinusoid in the steady state. As a result $u \in$ PE and Assumption 4 is satisfied.

From (55), it is clear that an accurate measuring of the input voltage and current is vital to achieve control objective. Hence, the state and parameter estimation given by (28), (34) and (35) can be used in conjunction with the aforementioned dynamic control law. Notice that, with respect to (14), singularity problem can be occurred when $\hat{E}$ is used to estimate $I_0$ in $i_d$. In order to circumvent the problem, we assume that $E^{-1}$ is lumped in the arbitrary gain $c$ in (54) and the signal

$$e := Ee, \quad (56)$$

is fed to the filter (54) instead of $e$. An estimate of $e$ can be obtained by

$$\hat{e}(t) = \hat{q}\cos(\omega t + \hat{\rho} - \hat{p}) - \hat{E}(uv - k\hat{i}), \quad (57)$$

with

$$\hat{q} := \sqrt{(\hat{E}^2 - 2kGV_d^2)^2 + (2GV_d^2 L\omega)^2}, \quad (58)$$

$$\hat{p} := \arctan\left(\frac{\hat{E}^2 - 2kGV_d^2}{-2GV_d^2 L\omega}\right). \quad (59)$$

Finally, the control system in the closed loop with the estimator will be complete by

$$\dot{u} = \frac{1}{v}\left(-\frac{u^2}{C}\hat{i} + \hat{w}\right), \quad (60)$$

with

$$\hat{w}(s) = d\frac{s^2 + as + b}{s^2 + \omega}\hat{e}(s), \quad (61)$$

where $d$ is related to $c$ in (54) with $d = c/E$.

*Remark 4:* Individually estimation of the AC-side parasitic resistance proposed in [6] is extra when an estimation of input voltage is making. Note that, its effect can be added in (55) as a small voltage drop in phase with the AC-source (see equation (23) in [6]). Therefore, the incorporation of this uncertainty into the estimated amplitude of input voltage is done in the modified control law above.

## V. SIMULATION RESULTS

The proposed estimator and controller are carried out on the system with the parameters given in Table I, via the Simulink® of MATLAB® R2016a. The system is modelled by Simscape™ Power Systems™ Simscape Component blocks as shown in Fig. 3. In order to take into account the practical restrictions, a current limiter block is considered in the model which prevents from drawing an instantaneous input current up to 14 A. The internal resistance of the AC-source is considered by $r$ in the model shown by Fig. 3. The parasitic parameters of the blocks are set to the default values and the switching frequency of the PWM is set to 10 KHz. Also, a block diagram of the control system is shown in Fig. 4. The controller and estimator gains are given in Table II. The values of the system and controller parameters are borrowed from the experimental setup in [6] for comparison purposes.

TABLE I. PARAMETERS VALUES OF THE SYSTEM

| Parameter | Value | Unit |
|---|---|---|
| $E$ | 150 | V |
| $\omega$ | $100\pi$ | rad/sec |
| $\rho$ | $2\pi/3$ | rad |
| $r$ | 2 | $\Omega$ |
| $L$ | 2.13 | mH |
| $C$ | 1100 | $\mu$F |
| $G$ | 1/87 | S |
| $V_d$ | 200 | V |

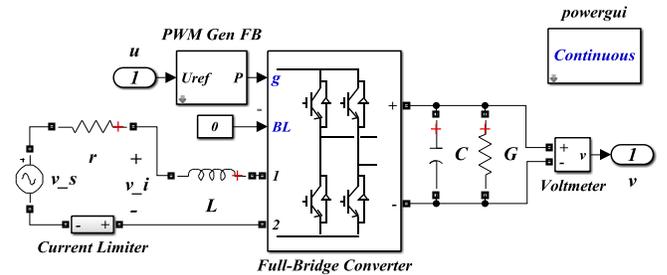

Figure 3. The interface between the blocks in the Simscape™ Power Systems™ Simscape Component.

Figure 4. Gains Values of the Estimator and Controller

| Estimator Gain | Value | Controller Gain | Value |
|---|---|---|---|
| $k$ | $2 \times 10^{-4}$ | $a$ | 1200 |
| $\mathcal{T}$ | $10 \times \mathcal{I}_2$ | $b$ | $2 \times 10^5$ |
| $\mathcal{D}$ | $20 \times \mathcal{I}_2$ | $d = c/E$ | 460/15 |
| | | $k$ | 15 |

The simulations intend to evaluate the behaviour of the proposed control system in confronting with step changes in amplitude and phase of AC-source, the desired output voltage, and load conductance. The evaluation is done by computing the DC error of the output voltage, and the harmonic characteristics of the input current such as total harmonic distortion (THD), displacement angle, and power factor, as shown in Fig. 5.

The performance of the dynamic controller (53)-(55) (so-called internal model (IM) controller [6]) is compared with the certainly equivalent version (57)-(61) shown in Fig. 4. All initial values of the controller and estimator are set to zero, whereas the system leaves the initial values $i(0) = 1.5$ A and $v(0) = 100$ V, to reach the set-point $V_d = 200$ V. The simulation results are shown in Table III and Fig. 6 and represents a significant error in the DC value of the output voltage whereas the error is compensated in the certainly equivalent version. The error can be imposed by the parasitic resistance in the circuit elements. The simulation results underscore the role of the adaptation in the certainly equivalent version of the controller in confronting with such uncertainties.

Fig .7 shows the response of the proposed estimator and controller to the amplitude change from 150 to 120 V at $t = 0.5$ sec and the phase change from $2\pi/3$ to 0 rad of the AC-source at $t = 0.9$ sec. Simulation results represent a fast convergence in the estimator and controller performance. Fig .7 shows that $\rho$ is estimated exactly by the estimator, whereas $\hat{E}$ tracks a new value to compensate the voltage drop induced by the parasitic resistance, such as $r$.

Fig. 8 shows the response of the control system to a desired output change from 200 to 160 V at $t = 1.3$ sec and conductance change from 87 to 51 S at $t = 1.7$ sec. Observer that the output reaches to its new reference without undershoot. Also, it remains close to the desired DC value for the significant inductance change. Fig. 8 (bottom graph) shows that, the current estimation tracks the actual value for the mentioned changes. Also, it can be seen that the estimator behaves in a robust manner for the inductance change.

## VI. CONCLUSION

In this paper the effect of the phase difference between the input voltage and current on the power quality is investigated. The mathematic analyses show that the big values of the phase shift let to increase the DC error and the amplitude of the harmonic in the output voltage. A new nonlinear, globally exponentially convergent, robust estimator is designed via I&I based filtered transformation. The input current and voltage are estimated from the output voltage via a fifth-dimensional estimator. Also, an application of the estimator is presented in conjunction with a well-known dynamic controller. Simulation results illustrate a fast convergence and robust performance of the proposed scheme.

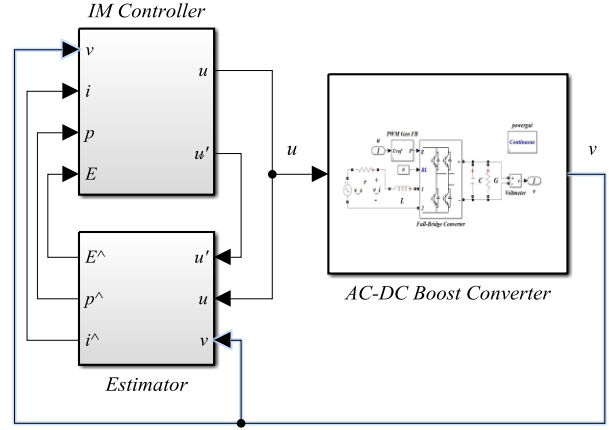

Figure 5. The structure of the control system in MATLAB Simulink.

TABLE II. HARMONIC ANALYSIS

| Controller | Displacement (deg) | THD (%) | Power Factror | DC Error (V) |
|---|---|---|---|---|
| IM | 0.0174 | 0.4135 | 0.9998 | 8.20 |
| IM + Estimator | 0.0065 | 2.1010 | 0.9998 | 0.5 |

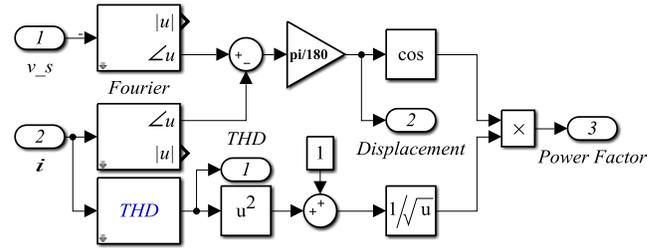

Figure 6. The structure of the harmonic characteristic computation by MATLAB Simulink.

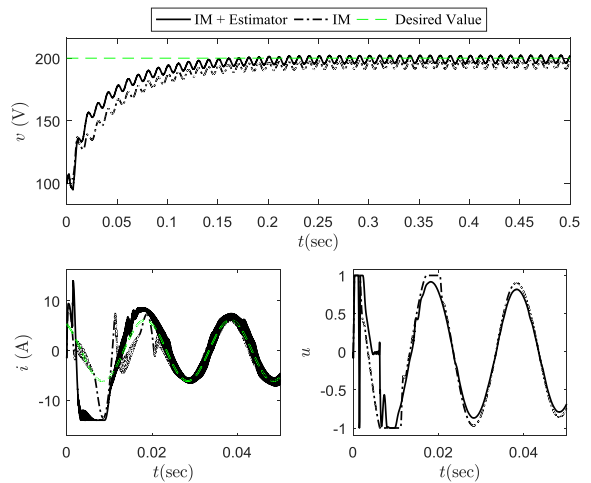

Figure 7. Thime histories of the states and control signal for the IM and IM+Estimator controller.

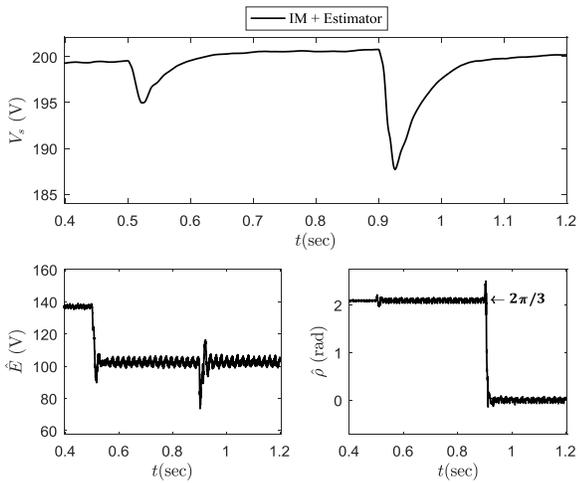

Figure 8. Response of the IM+Estimator controller to a amplitude and phase change in AC-source. Top graph: DC value of the output volatge. Bottom graph: estimated parameters.

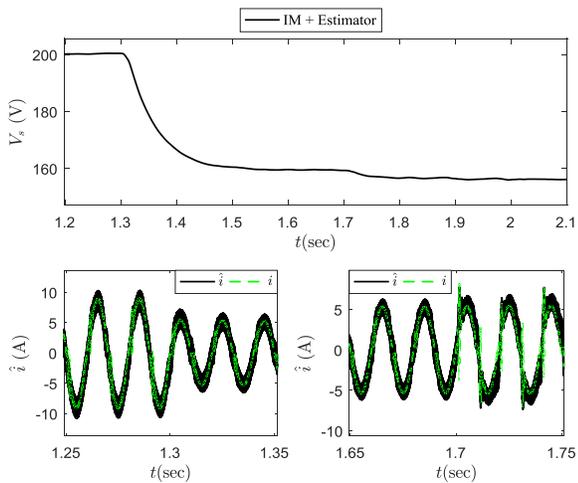

Figure 9. Response of the IM+Estimator controller to a voltage reference and conductance change. Top graph: DC value of the output volatge. Bottom graph: estimated state.